\documentclass[12pt]{article}
\begin{document}
\begin{titlepage}

\title{Dynamical calculation of $d^*$ mass and $NN$ decay width in the
quark delocalization, color screening model}

\author{
 Jialun Ping\thanks{\small \em E-mail: jlping@pine.njnu.edu.cn}\\ 
{\small \em Department of Physics, Nanjing Normal University,}\\
{\small \em  Nanjing, 210097, P.R. China}\\
Fan Wang\thanks{\small \em E-mail: fgwang@chenwang.nju.edu.cn}\\ 
{\small \em Center for Theoretical Physics and Department of Physics, }\\
{\small \em Nanjing University, Nanjing, 210093, P. R. China}\\
T. Goldman\thanks{\small \em E-mail: tgoldman@lanl.gov}\\
{\small \em Theoretical Division, Los Alamos National Laboratory, }\\
{\small \em Los Alamos, NM 87545, USA}
}

\date{July 28, 2000}
\maketitle

\begin{abstract}
The mass estimate of the $d^* (IJ^P=03^+)$ dibaryon is improved by a
dynamical calculation in the quark delocalization, color screening
model. The partial decay width of $d^*$ into an $NN$ D-wave state is
also obtained.  The mass obtained is slightly larger than that obtained
in adiabatic calculations, due to the anharmonicity of the effective
potential between two $\Delta$'s. The value of the width obtained due
to tensor one-gluon-exchange is about 5 MeV, comparable in
magnitude to earlier results found using pion exchange.

{\noindent 12.39.-x, 14.20.Pt, 13.75.Cs}
\end{abstract}

\vspace*{-8.0in}
\begin{flushright}LA-UR-00-3209\\
\vspace*{-0.05in}
nucl-th/0007040
\end{flushright}
\vspace*{8.0in}

\end{titlepage}

\section{Introduction} 
Quantum Chromodynamics (QCD) is widely accepted as the fundamental
theory of the strong interaction. However, the low energy behavior of
QCD remains a challenge to understand. Lattice QCD can calculate the
structure of simple quark-gluon systems but it remains extremely
difficult to calculate complicated systems such as two-hadron
interactions and the structure of multiquark systems.  Quark models
were inspirational in the development of QCD and may play a similar
role in understanding the low energy behavior of QCD.

The nonrelativistic constituent quark model has been quite successful
in understanding hadron spectroscopy~\cite{Isgur}. This model has been
extended by many groups to the description of baryon-baryon (B-B)
interactions. There is a broad consensus that the repulsive core of the
N-N interaction can be attributed to the quark internal structure of
the nucleon. In this respect, the repulsive core is quite similar to
that of molecular forces originating from the electronic internal
structure of atoms. However, the intermediate range attraction of the
N-N interaction remains absent in many quark model approaches and a
scalar meson exchange has frequently been invoked to remedy this.

Spontaneous chiral symmetry breaking supports the view that Goldstone
boson exchanges occur even between constituent quarks. Based on this,
recently a phenomenological quark model has been proposed by Glozman
{\em et al.}~\cite{Glozman}. The model gives a rather good description
of the baryon spectrum and it has also been applied to NN interactions
and dibaryons~\cite{Stancu}.But a question remains as to what extent
this picture should be used: Should it be used for long range pion
exchange only, or extended to intermediate two pion or $\sigma$
exchange or even to include short range vector and other heavy meson
exchange? At what point should gluon exchange between quarks be
replaced? Of course, a great deal of further study is needed to settle
such questions definitively.

One point we would like to emphasize here is that such an approach
would qualitatively differentiate between nuclear and molecular forces,
which have been known for more than half a century to be similar in
many features, the energy and length scale differences
notwithstanding~\cite{bohr}.  Recently, Anderson has pointed out that
it is questionable to attribute the N-N interaction to meson
exchange~\cite{anderson} except for the longest range portion of pion
exchange. $NN$ partial wave analysis can only justify long range pion
exchange but not other, heavier meson exchanges~\cite{Timmermans}.

In the quark model approaches mentioned above, a two-body confinement
potential is usually used. We recognize that this is a highly
simplified model assumption. Quark confinement is a nonperturbative
property of low energy QCD. There might be important nonperturbative
features missed in every two-body confinement potential model.  For
example, three-gluon exchange between three quarks and three-body
instanton interactions do not contribute within a colorless meson or
baryon, but do contribute to a multiquark system~\cite{lsg,met}.  There
are many other types of quark interactions which result from multigluon
exchanges which can not be included in a two body confinement
potential.  The QCD condensates within a nucleon and between two
nucleons might be different. These arguments show that a direct
extension of the two body confinement to the multiquark system has not
been justified even though it might be a good approximation in the
single hadron case.  The absence of experimentally observed color Van
der Waals forces has been a problem for the two body confinement
potential model for a long time.  Multiquark systems provide more
variations of low energy behavior of QCD and allow for further tests of
the phenomenology built into quark models, especially the nature of
confinement.

The quark delocalization and color screening model (QDCSM) has been
developed in which an unusual confinement parameterization has been
introduced. Following the concept of electron delocalization in
molecular orbitals, quark delocalization is introduced to enlarge the
variational Hilbert space to include various deformed six-quark
configurations. The usual pair of three-quark clusters and six-quark,
"bag", configurations are the two extremes of this space~\cite{prl}.

With only one adjustable parameter, the color screening constant $\mu$,
the existing $NN$ ($IJ = 01, 10, 11, 00$), $N\Lambda$ ($IJ = \frac{1}{2}0,
\frac{1}{2}1$) and $N\Sigma$ ($IJ = \frac{1}{2}0, \frac{1}{2}1,
\frac{3}{2}0, \frac{3}{2}1$) scattering data have all been fit
qualitatively~\cite{prl,prc53,moda10,npa673} within this model. A
relativistic version of this model has recovered $^4He$, $^3He$, $^3H$
as approximate four- and three-nucleon systems starting from
appropriate 12- and 9-quark configurations~\cite{npa481,H3,plb324}.

Both the relativistic and the nonrelativistic version give almost the
same dibaryon spectrum, reproducing small deuteron binding and the zero
energy $NN$ $IJ = 10$ resonance. An $IJ^{P} = 03^+$ state, called the
$d^*$, was found to be most interesting because of its large binding
energy ($\sim$350~MeV) and small decay width
($\sim$1~MeV)~\cite{prl,Goldman,wpg,moda13,CWW}.  This state appears in
and has been studied in many models.  Kamae and Fujita obtained almost
the same large binding~\cite{Kamae}. Other models obtained smaller
binding~\cite{Mulders,Yazaki,Walet}.

The experimental situation regarding dibaryons remains unsettled.
Precise $np$ total cross section measurements provide a stringent limit
on the $NN$ decay width of any dibaryon in the $d^*$ mass
range~\cite{Lisowski}. Similarly, the $H$ particle has been studied in
many models and searched for experimentally for more than 20 years
without any indication of its existence~\cite{Jaffe}. There is an
$IJ^{P} = 00^-$ state, $d'$, which seems to have experimental
support~\cite{Bilger} but which is hard to accommodate in a six-quark
model space~\cite{Buchmann,Ping}. Newer experiments using simple
systems have not confirmed the existence of a $d'$
signal~\cite{ppexp}.  A high-mass dibaryon state predicted by Lomon
{\em et al}~\cite{Lomon} finds support from SATURNE $pp$ scattering
data~\cite{Lehar}.

This paper reports further study of the $d^*$ binding and decay width
with the aim of providing more reliable information for experimental
searches for this dibaryon.  The paper is organized as follows: A
brief review of the features of the QDCSM appears in Sect.~2. 
Our calculation method is presented in Sect.~3. Results and discussion 
are provided in Sect.~4. The last section gives a summary.

\section{Model Hamiltonian and wave function} 
The details of the QDCSM can be found in Ref.\cite{prl,wpg}. Here only   
the model Hamiltonian and wavefunction used in our calculations are  
repeated. 

The Hamiltonian for the 3-quark system is the same as the usual potential
model~\cite{Isgur}, and for a 6-quark system, it is assumed to be
\begin{eqnarray} 
H_6 & = & \sum_{i=1}^6 (m_i+\frac{p_i^2}{2m_i})-T_{CM} +\sum_{i<j=1}^{6} 
    \left( V_{ij}^C + V_{ij}^G \right) ,   \nonumber \\ 
V_{ij}^G & = & \alpha_s \frac{\vec{\lambda}_i \cdot \vec{\lambda}_j }{4} 
 \left[ \frac{1}{r}-\frac{\pi \delta (\vec{r})}{m_i m_j} 
 \left( 1+\frac{2}{3} \vec{\sigma}_i \cdot \vec{\sigma}_j \right)
 + V_T \right], 
   \label{hamiltonian} \\
V_T & = & \frac{1}{4m_im_j} \left[ \frac{3(\vec{\sigma}_i \cdot \vec{r})
  (\vec{\sigma}_j \cdot \vec{r})}{r^5} - \frac{\vec{\sigma}_i \cdot 
  \vec{\sigma}_j}{r^3} \right], \nonumber \\
V_{ij}^C & = & -a_c \vec{\lambda}_i \cdot \vec{\lambda}_j 
\left\{ \begin{array}{ll} 
 r^2 & 
 \qquad \mbox{if }i,j\mbox{ occur in the same baryon orbit}, \\ 
 \frac{1 - e^{-\mu r^2} }{\mu} & \qquad
 \mbox{if }i,j\mbox{ occur in different baryon orbits}, 
 \end{array} \right. \nonumber
\end{eqnarray} 
where $\vec{r}=\vec{r_i}-\vec{r_j}$ and 
all other symbols have their usual meaning except the confinement
potential $V^C_{ij}$ which will be explained below. 

It is well known that confinement is a nonperturbative QCD effect; in
general one does not expect it to be described by a sum of two-body
interactions. We model confinement as follows: In our approach, the
fractional parentage expansion method is used and the matrix elements
of the six-quark Hamiltonian are simplified to a four-body overlap and
a two-body matrix element. We assume the following recipe to determine
the two body matrix element of confinement: The interaction takes the
normal, unscreened form (quadratic in $r$) when the interacting quarks
always remain in the same baryon orbit $\phi$ (see Eq.(9) below), both
before and after interaction; otherwise the interaction takes the
screening form (second form shown in the last of
Eqs.(\ref{hamiltonian})). Although this has not been demonstrated to be
correct, it is more sophisticated than the usual, simple two-body
confining interaction, and does include a physically reasonable model
of non-local, nonperturbative effects.

Even though we use a two-body interaction form to evaluate the matrix
elements, our model is, nonetheless, not a potential model. It is an
effective matrix element approach extended from bound states to
scattering states. It does reduce to the usual two-body confinement
interaction within a single hadron. It also has the usual meaning of a
two-body interaction in the asymptotic regime although not in
intermediate regions. The main physics introduced is the recognition
that the confining interaction between two nucleons might be different
from that within a nucleon. In particular, we represent the nonlocal,
nonperturbative backflow of color (and pair creation, as seen in
lattice studies~\cite{lats}) by screening of the confining potential.
If this model assumption of the confinement is not a good
representation of the physics involved, we would expect that to appear
in a disagreement between our calculational results and data. The
quality of the fit to B-B scattering data referred to in the
introduction shows that there is no obvious contradiction. Our
predictions with respect to dibaryon states will provide a further
test.

We use the resonating group method (RGM) to carry out a dynamical
calculation.  Following the nomenclature of Ref.\cite{Buchmann}, we
write the conventional ansatz for the two-cluster wavefunction as
\begin{equation} 
|\Psi_{6q}\rangle = {\cal A} \sum_{L} \left[ 
  [\Phi_{B_1}\Psi_{B_2}]^{[\sigma]IS} \otimes \chi_{L}
(\vec{R}) \right]^J,
\label{wf6q} 
\end{equation} 
where $[\sigma]=[222]$ gives the total color symmetry and all other
symbols have their usual meanings. $\Psi_{B_i}$ is the 3-quark cluster
wavefunction (after removal of the center of mass motion),
\begin{equation} 
\Psi_{B_i} = \left( \frac{2}{3\pi b^2} \right)^{3/4}
             \left( \frac{2}{4\pi b^2} \right)^{3/4} 
  e^{-\left( \frac{\vec{\lambda}_i^2}{3b^2}
+\frac{\vec{\rho}_i^2}{4b^2}\right)}
  \eta_{I_iS_i}(B_i)\chi_c(B_i), 
\label{wf3q1} 
\end{equation} 
where $\chi_c(B_i)$ is the internal color wavefunction of the baryon and
the Jacobi coordinates are defined as follows,
\begin{eqnarray} 
\vec{\rho}_1 & = & \vec{r}_1 - \vec{r}_2, ~~~~~~~~~~~~~~~~~~
 \vec{\rho}_2  =  \vec{r}_4 - \vec{r}_5, \nonumber \\
\vec{\lambda}_1 & = & \vec{r}_3 - \frac{1}{2} (\vec{r}_1+\vec{r}_2), ~~~~~~~
 \vec{\lambda}_2  =  \vec{r}_6 - \frac{1}{2} (\vec{r}_4+\vec{r}_5), \\ \nonumber
\vec{R}_{B_1} & = & \frac{1}{3} (\vec{r}_1 + \vec{r}_2+\vec{r}_3), ~~~~~
 \vec{R}_{B_2} = \frac{1}{3} (\vec{r}_4 + \vec{r}_5+\vec{r}_6), \\
\vec{R} & = & \vec{R}_{B_1} - \vec{R}_{B_2},  ~~~~~~~~~~~
 \vec{R}_C = \frac{1}{2} (\vec{R}_{B_1} + \vec{R}_{B_2}). 
   \label{jacobi} \nonumber
\end{eqnarray} 
From the variational principle, after variation with respect to the relative
motion wavefunction $\chi (\vec{R})= \sum_L \chi_{L} (\vec{R})$, one
obtains the RGM equation
\begin{equation}
\int H(\vec{R},\vec{R}')\chi (\vec{R}') d\vec{R}' =
  E \int N(\vec{R},\vec{R}')\chi (\vec{R}') d\vec{R}', \label{RGM}
\end{equation}
where $ H(\vec{R},\vec{R}'), N(\vec{R},\vec{R}')$ are Hamiltonian and
norm kernels, respectively. Their detailed expressions can be found in
Ref.\cite{Buchmann}.

The energies, $E$, and the wavefunctions, $\chi (\vec{R})$, are
obtained by solving the RGM equation. In practice, it is not convenient
to work with the RGM expressions. We introduce generator coordinates,
$\vec{S}_i$, to expand the relative motion wavefunction, $\chi
(\vec{R})$,
\begin{equation}
\chi (\vec{R}) = \frac{1}{\sqrt{4\pi}} \sum_{L} \left( \frac{3}{2\pi b^2}
  \right)^{3/4} \sum_i C_{i,L} \int e^{-\frac{3}{4b^2} (\vec{R}-\vec{S}_i)^2}
  Y^L (\hat{\vec{S}}_i) d\Omega_{S_i},
\end{equation}
After the inclusion of the center of mass motion,
\begin{equation}
\Phi_C (\vec{R}_C) = \left( \frac{6}{\pi b^2} \right)^{3/4} e^{-\frac{3}{b^2}
  \vec{R}_C^2},
\end{equation}
the ansatz, Eq.(\ref{wf6q}), can be rewritten as
\begin{eqnarray}
\Psi_{6q} & = & {\cal A} \sum_{i,L} C_{i,L} \int 
  \frac{d\Omega_{S_i}}{\sqrt{4\pi}}
  \prod_{\alpha=1}^{3} \phi_{\alpha} (\vec{S}_i) 
  \prod_{\beta=4}^{6} \phi_{\beta} (-\vec{S}_i)  \nonumber \\
  & &  \left[ [\eta_{I_1S_1}(B_1)\eta_{I_2S_2}(B_2)]^{IS}
  Y^L(\hat{\vec{S}}_i) \right]^J [\chi_c(B_1)\chi_c(B_2)]^{[\sigma]}
	\label{wf6qn}
\end{eqnarray}
where $\phi_{\alpha}(\vec{S}_i), \phi_{\beta}(-\vec{S}_i)$ are the single 
particle orbital wavefunctions with different reference centers,
\begin{eqnarray}
\phi_{\alpha}(\vec{S}_i) & = & \left( \frac{1}{\pi b^2} \right)^{3/4}
   e^{-\frac{1}{2b^2} (\vec{r}_{\alpha} - \vec{S}_i/2)^2} \nonumber \\
\phi_{\beta}(-\vec{S}_i) & = & \left( \frac{1}{\pi b^2} \right)^{3/4}
   e^{-\frac{1}{2b^2} (\vec{r}_{\beta} + \vec{S}_i/2)^2}. \label{1q}
\end{eqnarray}
With the reformulated ansatz, Eq.(\ref{wf6qn}), the RGM equation
Eq.(\ref{RGM}) becomes an algebraic eigenvalue equation,
\begin{equation}
\sum_{j,L} C_{j,L} H^{L,L'}_{i,j} = E \sum_{j} C_{j,L'} N^{L'}_{i,j}
   \label{GCM}
\end{equation}
where $N^{L'}_{i,j}, H^{L,L'}_{i,j}$ are the Eq.(\ref{wf6qn})
wavefunction overlaps and Hamiltonian matrix elements (without the
summation over $L'$), respectively. By solving the generalized eigen
problem, we obtain the energies of 6-quark system and corresponding wavefunctions.

In the QDCSM, the single particle orbital wavefunctions are
delocalized.  To implement this here, we modify Eqs.(\ref{1q}) as
follows:
\begin{eqnarray}
\phi_{\alpha}(\vec{S}_i) & \rightarrow & \psi_{\alpha}(\vec{S}_i ,
\epsilon) = \left( \phi_{\alpha}(\vec{S}_i) + \epsilon \phi_{\alpha}
(-\vec{S}_i)\right) /N(\epsilon), \nonumber \\
\phi_{\beta}(\vec{S}_i) & \rightarrow & \psi_{\beta}(-\vec{S}_i ,
\epsilon) = \left(\phi_{\beta}(-\vec{S}_i) + \epsilon \phi_{\beta}
(\vec{S}_i)\right) /N(\epsilon), \\
& & N(\epsilon) = \sqrt{1+\epsilon^2+2\epsilon e^{-S_i^2/4b^2}}. \nonumber
\end{eqnarray}
 
It is straightforward to extend the method to multichannel coupling. 
In the multichannel case, the ansatz used is
\begin{eqnarray}
\Psi_{6q} & = & {\cal A} \sum_{k} \sum_{i,L_k} C_{k,i,L_k} \int 
  \frac{d\Omega_{S_i}}{\sqrt{4\pi}}
  \prod_{\alpha=1}^{3} \psi_{\alpha} (\vec{S}_i , \epsilon) 
  \prod_{\beta=4}^{6} \psi_{\beta} (-\vec{S}_i , \epsilon)  \nonumber \\
  & &  \left[ [\eta_{I_{1k}S_{1k}}(B_{1k})\eta_{I_{2k}S_{2k}}(B_{2k})]^{IS_k}
  Y^{L_k}(\hat{\vec{S}}_i) \right]^J [\chi_c(B_1)\chi_c(B_2)]^{[\sigma]}
	\label{mutli}
\end{eqnarray}
where $k$ is the channel index.  The eigen equation is similar to
Eq.(\ref{GCM}), with an additional summation over $k$.  For example,
for $IJ=03$, we have $k=1, 2, 3, 4$, corresponding to the channels
$\Delta\Delta~ S=3~L=0$, $\Delta\Delta~ S=3~L=2$, $NN~ S=1~L=2$, and
$\Delta\Delta~ S=1~L=2$.

\section{Calculation method} 
 
To further simplify the calculation of the matrix elements of the
six-quark Hamiltonian, the physical bases, Eq.(\ref{wf6qn}), are first
expanded in terms of symmetry bases (group chain classification bases).
Then the powerful fractional parentage expansion method is used to
calculate the matrix elements between symmetry bases of the six-quark
system. Finally, the matrix elements between physical bases are
obtained by the transformation between the physical bases and symmetry
bases. The details can be found in Ref.\cite{chen,wpg2}.

The partial width for $d^*$ decay into the $NN$ D-wave state is
obtained using ``Fermi's Golden Rule'',
\begin{eqnarray}
\Gamma & = & \frac{1}{7} \sum_{M_{J_i}, M_{J_f}}
  \frac{1}{(2\pi)^2} \int p^2 dp~d\Omega~ \delta(E_f-E_i) |M|^2
  \nonumber \\
  & = & \frac{1}{7} \sum_{M_{J_i}, M_{J_f}}
  \frac{1}{32\pi^2} m_{d^*} \sqrt{m_{d^*}^2 - 4 m_N^2} \int |M|^2
  d\Omega,
  \label {width}
\end{eqnarray}
where $M$ denotes the nonrelativistic transition matrix element, and
$M_{J_i}$ and $M_{J_f}$ are the spin projections of the initial and
final states.
\begin{equation}
M = \langle d^* | H_I | [\Psi_{N_1} \Psi_{N_2}]^{IS} e^{i \vec{p}\cdot \vec{R}}
 \rangle, \label{M}
\end{equation}
where $\vec{R}$ is as above and $p = \frac{1}{2}\sqrt{m_{d^*}^2 - 4
m_N^2}$ is the available relative momentum between the nucleons as
determined by the energy conserving $\delta$-function in
Eq.(\ref{width}). The interaction Hamiltonian, $H_I$, here is $H_I =
V_T$.

By expanding the plane wave in terms of spherical harmonics, and taking
into account angular momentum conservation, the transition matrix
element can be put in the form,
\begin{equation}
M = -4\pi \left\langle d^* |H_I| C^{JM_J}_{SM_S,LM_L}
   \left[ [\Psi_{N_1}\Psi_{N_2}]^{IS}
   Y^L(\hat{\vec{R}})(Y^{L}(\hat{\vec{p}}))^* \right]^J j_{L}(pR+\delta_L) 
  \right\rangle. \label{mtxelwvfcn}
\end{equation}
where we have included the phase shift, $\delta_L$,  induced by the
interaction in the spherical Bessel function. Of course, the phase
shift is only apparent at large separations.

On the other hand, the relative motion wavefunction between two
nucleons can be determined from either the RGM equation or the
algebraic equation, Eq.(\ref{GCM}). The relative motion wavefunction
can be written as
\begin{equation}
\chi_L (\vec{R}) = \sqrt{4\pi} \sum_i C_{iL} 
 \left( \frac{3}{2\pi b^2}\right)^{3/4}
  e^{-\frac{3}{4b^2} (\vec{R}^2+\vec{S}_i^2)} i_L(\frac{3}{2b^2}RS_i)
  Y^L(\hat{\vec{R}}),
\end{equation}
where $i_L(z)$ is the modified spherical Bessel function.
If we can match the expansion
\begin{equation}
\sum_{i} C_{iL} \left( \frac{3}{2\pi b^2}\right)^{3/4}
  e^{-\frac{3}{4b^2} (\vec{R}^2+\vec{S}_i^2)} i_L(\frac{3}{2b^2}RS_i)
  \label{match}
\end{equation}
to the function $j_L(pR+\delta_L)$ for the separation in the calculated
region, then the wavefunction used in Eq.(\ref{mtxelwvfcn}) is just the
wavefunction defined in Eq.(\ref{wf6qn}) with an additional factor of
$\sqrt{4\pi} (Y^L(\hat{\vec{p}}))^*$, which contributes a factor $4\pi$
to the decay width after the integration over the angular space,
$\Omega$.

We have used a type of box normalization method to calculate the
transition matrix element.  Because the wavefunction of the initial
bound state $d^*$ is compact, its amplitude at large separation ($R >
R_0\sim 3.3$ fm) can be safely neglected. By contrast, the final $NN$
D-wave state with the same energy as that of the $d^*$ is a scattering
state, so it extends over the entire space.  However, the transition
matrix element depends only on the overlap between the $d^*$ and $NN$
D-wave states, so only that part of the wavefunction of the $NN$ D-wave
with separation less than $R_0$ fm can contribute. To determine the
wavefunction of the final state $NN$ D-wave in this finite range, the
RGM equation is used so that the final state interaction is taken into
account automatically.  By adjusting the box size, the correct internal
relative momentum $NN$ D-wave state can be obtained to meet the
requirement of energy conservation in the decay. Finally, a
normalization correction due to the finite range of the wavefunction is
needed, which is obtained by matching the expression Eq.(\ref{match})
with $j_L(pR+\delta_L)$. In this way, we are able to address the bound
state problem and the scattering state problem on the same footing. For
example, the phase shifts of the scattering state can be obtained from
the comparison of the calculated wavefunction with the free Bessel
function, $j_L(pR)$.

\section{Results and discussion}
As a test of the model, we first attempted a calculation of the
deuteron.  The parameters we used in the calculation, which are fixed
by baryon properties and $NN$ scattering, are
$$
m_u=m_d=313~ \mbox{MeV}, ~b=0.603~ \mbox{fm}, ~a=25.13~ 
\mbox{MeV/fm$^2$}, ~\alpha_s=1.54, ~\mu=1.0~\mbox{fm}^{-2},
$$
These parameter values are very similar to those usually appearing in
constituent quark models~\cite{Isgur}.  The single channel and channel
coupling calculations do show an attraction between nucleons, but it is
insufficient to form a bound state. The mixing of the D-wave into the
deuteron state is also too small (less than 1\%) to affect the
result.  The deuteron can be forced to form in the QDCSM by increasing
the color screening parameter. However the small radius of the
resulting object and negligible D-wave mixing demonstrate that it is
not physically correct to do this.

Together with the absence of a long range tail for the $NN$
interaction, this serves to emphasize the need for the long range part
of one pion exchange (OPE), which is absent from our model.  To
reproduce the deuteron, which is an extremely extended object, it will
be necessary to extend the QDCSM to include OPE.  To avoid double
counting, a short-distance cutoff of the pion-quark coupling is also
required. Preliminary calculations show that the deuteron can be well
reproduced in this manner.  We will address these issues in a future
publication~\cite{nxtppr}.

The case of the $d^*$ is quite different, however. The small size and
the large delocalization of quarks shows that it is a highly compact
object. The effect of OPE should be considerably reduced compared to
the case of the deuteron.  Using the same parameters, a dynamical
calculation of the $d^*$ was carried out.  These results appear in
Table I. We find that the mass of $d^*$ is somewhat larger than that
obtained by our earlier adiabatic calculation~\cite{wpg}, mainly
because the relative motion energy found is larger than that of the
zero-point oscillation energy used in the adiabatic calculation.
Although this $d^*$ mass exceeds the $NN\pi\pi$ threshold, it should
remain a narrow resonance due to the extremely small phase space
available~\cite{CWW1}.

\begin{center}
Table I. Mass and radius of the $d^*$ in the QDCSM. \\ 
sc= single channel; cc2 = $\Delta\Delta$ S-wave + $NN$ D-wave; 
cc4 = full four-channel coupling.

\begin{tabular}{c|cccc} \hline
 & sc  & cc2 & cc4 & adiabatic~\cite{wpg} \\ \hline
mass (MeV) & ~~~2186~~~ & ~~~2180~~~ & ~~~2176~~~ & ~~~2134~~~  \\
~~~$\sqrt{<r^2>}$ (fm)~~~ & 1.2 & 1.2 & 1.3 & 1.4 \\ \hline
\end{tabular}
\end{center}

We have also calculated the partial decay width of $d^*$ into the $NN$
D-wave, using the method introduced in last section. The result is
dominated by the stretched ($m_J = \pm 3$) states which contribute
$\Gamma^{\mbox{str.}} = 1.95$ MeV each to the sum over spin
projections. We find $\Gamma = 4.32$ MeV for the total width. This
result is similar to that obtained from meson exchange~\cite{CWW}. It
should be expected that the total width will be increased by including
OPE as discussed above for the deuteron. Although it is generally
believed that the tensor interaction due to OPE is much stronger than
that from one gluon exchange, we expect this increase will be limited
due to the combination of the compactness of the $d^*$ and the required
short-distance cutoff of the pion-quark coupling.

\section{Summary}
In the framework of QDCSM, dynamical calculations of the $d^*$ dibaryon
and the partial decay width of the $d^*$ into an $NN$ D-wave state were
carried out.  The results include a slight increase in the mass of the
$d^*$ over the adiabatic result, rising above the $NN\pi\pi$
threshold.  However, the small phase space still suggests that the
resonance will be narrow and, indeed, the calculated $d^* \rightarrow
NN$ D-wave decay width is on the order of a few MeV.

Our calculation of the deuteron shows the predictive power of the
QDCSM  on the one hand, but on the other, that OPE must be included in
the calculation of extended objects (with small delocalization).
However, we do not expect that the addition of OPE will have a large
effect on the calculation of $d^*$, because of its compactness.
Addition of OPE to the QDCSM is underway and preliminary results show
that both the D-wave component and the radius of the deuteron are
indeed improved. Details will be presented in a forthcoming
paper~\cite{nxtppr}.

The calculated $d^* \rightarrow NN$ D-wave decay width is larger than
allowed by the Lisowski~\cite{Lisowski} data. However for a state as
high as 2.2 GeV, sea quark excitation or the $NN\pi$ component may well
be important. Inclusion of this component may affect the mass and decay
width strongly. In particular, the $d^* \rightarrow NN\pi$ decay width
may be significantly increased~\cite{CWW1}. We believe that it is still
too early to conclude that the $d^*$ has been ruled out by precise $np$
total cross section measurements and that further calculations
including these complications are needed.

\vspace{0.2in}
This research is supported by the National Science Foundation of China,
the Fok Yingdung Educational Fund, the Natural Science Foundation of
Jiangsu Province and the U.S. Department of Energy under contract
W-7405-ENG-36.

\end{document}